\documentclass[amsmath,nofootinbib,twocolumn]{revtex4-1}
\usepackage{aas_macros,bm,graphicx}
\usepackage[colorlinks]{hyperref}
\pdfoutput=1

\let\originalleft\left
\let\originalright\right
\renewcommand{\left}{\mathopen{}\mathclose\bgroup\originalleft}
\renewcommand{\right}{\aftergroup\egroup\originalright}
\mathcode`\*="8000
{\catcode`\*=\active\gdef*{\mathclose{}\,\mathopen{}}}

\newcommand{\br}[1]{\left[#1\right]}
\newcommand{\cu}[1]{\left\{#1\right\}}
\newcommand{\pa}[1]{\left(#1\right)}
\newcommand{\ed}{\mathop{}\!\mathrm{d}}
\newcommand{\pd}{\mathop{}\!\partial}
\newcommand{\C}{\mathcal{C}}
\newcommand{\I}{\mathcal{I}}
\renewcommand{\O}{\mathcal{O}}

\begin{document}

\title{Inclined Pulsar Magnetospheres in General Relativity:\\
Polar Caps for the Dipole, Quadrudipole and Beyond}

\author{Samuel E. Gralla} 
\affiliation{Department of Physics, University of Arizona, Tucson, AZ 85721, USA}
\author{Alexandru Lupsasca}
\affiliation{Center for the Fundamental Laws of Nature, Harvard University, Cambridge, MA 02138, USA}
\author{Alexander Philippov}
\affiliation{Department of Astrophysical Sciences, Princeton University, Princeton, NJ 08540, USA}

\begin{abstract}
    In the canonical model of a pulsar, rotational energy is transmitted through the surrounding plasma via two electrical circuits, each connecting to the star over a small region known as a ``polar cap.''  For a dipole-magnetized star, the polar caps coincide with the magnetic poles (hence the name), but in general, they can occur at any place and take any shape.  In light of their crucial importance to most models of pulsar emission (from radio to X-ray to wind), we develop a general technique for determining polar cap properties.  We consider a perfectly conducting star surrounded by a force-free magnetosphere and include the effects of general relativity.  Using a combined numerical-analytical technique that leverages the rotation rate as a small parameter, we derive a general analytic formula for the polar cap shape and charge-current distribution as a function of the stellar mass, radius, rotation rate, moment of inertia, and magnetic field.  We present results for dipole and quadrudipole fields (superposed dipole and quadrupole) inclined relative to the axis of rotation.  The inclined dipole polar cap results are the first to include general relativity, and they confirm its essential role in the pulsar problem.  The quadrudipole pulsar illustrates the phenomenon of thin annular polar caps.  More generally, our method lays a foundation for detailed modeling of pulsar emission with realistic magnetic fields.
\end{abstract}

\maketitle

\section{Introduction}

Fifty years after the basic elements of pulsar theory were established \cite{Goldreich69}, self-consistent modeling remains challenging.  To simplify the problem, the community has mainly focused on a canonical choice of magnetic field: the pure dipole.  This endeavor has come to fruition in the last decade, as the dipole pulsar has been self-consistently modeled using increasingly detailed descriptions of the surrounding plasma, from force-free electrodynamics \cite{Contopoulos99, Gruzinov05, McKinney06, Timokhin06, Spitkovsky06, Kalapotharakos09, Li11, Kalapotharakos12, Petri12, Lehner12, Palenzuela13, Ruiz14, Petri16} to magnetohydrodynamics \cite{Komissarov06, Tchekhovskoy13} to kinetic theory \cite{Philippov14, Chen14, Belyaev15, Cerutti15, Philippov15a, Philippov15b}.

In contrast, alternative magnetic field configurations have remained relatively unexplored.  This is due mainly to the high cost of numerical simulations and the lack of an obvious alternative field configuration to choose.  However, there is little reason to believe in a pure dipole field, and the wide variation in emission properties among the known pulsars seems to be naturally accounted for by a correspondingly wide variation in the stellar magnetic field. It is therefore of interest to develop efficient techniques for exploring the effects of more realistic magnetic fields.

Building on our recent work in the axisymmetric case \cite{Gralla16b} (hereafter Paper I; see also Ref.~\cite{Belyaev16}), in this paper, we introduce a general method for determining the near-field charge and current flow (i.e., the pulsar polar caps) for a given magnetic field geometry on a given general relativistic star.  The key observation underpinning our analysis is that the pulsar problem contains a small parameter: the ratio of the stellar radius $R_{\star}$ to the light cylinder radius $R_L=c/\Omega$ (where $\Omega$ denotes the rotation rate and $c$ is the speed of light).  This parameter $\epsilon=R_\star/R_L$ is proportional to the surface rotation velocity and ranges from $10^{-4}$ to $10^{-1}$ for rotation-powered pulsars.

The small value of $\epsilon$ makes numerical work challenging, since both scales $R_\star$ and $R_L$ must be resolved.  Indeed, most numerical simulations are run at large values $\epsilon\sim1/5$ to reduce the dynamic range.  In contrast, analytic methods can shine at small values $\epsilon\ll1$.  To resolve both scales, one can use the method of matched asymptotic expansions \cite{Gralla16b}.  This involves making separate far $(r\gg R_\star)$ and near $(r\ll R_L)$ expansions, and matching in the overlap region $R_\star\ll r\ll R_L$ of shared validity.

We apply this method to a force-free magnetosphere surrounding a perfectly conducting general relativistic star \cite{Thorne68}.  In the far region, the equations reduce to the force-free magnetosphere of a rotating point dipole in flat spacetime, which we solve numerically.  In the near region, the equations become those of a static vacuum magnetic field in the Schwarzschild spacetime (representing the intrinsic magnetic field of the star), whose general solution is known in closed form.  Our near-far matching is expedited by a conserved quantity on magnetic field lines that generalizes the familiar field-aligned current (see App.~\ref{app:ConservedQuantity}).

The main quantities of interest are the leading-order charge and current near the star.  The charge is essentially induced by rotation, and is easily computed from the stellar magnetic field and rotation rate.  For the current, we fit for the conserved quantity in the far zone and then ``paint'' the associated current onto each field line, following it down to the surface of the star.  This last step amounts to finding Euler potentials for the stellar magnetic field, and hence is highly non-trivial in general.  It is, however, trivial to find Euler potentials for fields that are axisymmetric about some axis (not necessarily aligned with the rotation axis).  We thereby provide definite analytical formulae for the polar cap structure of an inclined pulsar with an arbitrary axisymmetric magnetic field.  For intrinsically nonaxisymmetric fields, the Euler potentials must in general be found numerically.  However, this task is still considerably simpler than running a complete self-consistent simulation.

While the magnetization is the most important, all of the stellar parameters affect the polar cap shapes and properties.  In particular, the compactness and moment of inertia control the redshift and frame-drag effects (respectively) of general relativity, and may be used to quantify the importance of these effects.  We emphasize, however, that in our approach, general relativistic effects are not added in piecemeal or ``by hand''; we simply begin with force-free electrodynamics in the spacetime of a rotating, conducting star and compute self-consistently to leading order in the rotation rate $\epsilon$.  In this limit, the magnetosphere is completely described by the stellar mass, radius, moment of inertia, rotation rate, and magnetization, all of which may be independently chosen.

We apply our method to the inclined dipole as well as to the inclined ``quadrudipole,'' a superposition of dipole and quadrupole fields \cite{Barnard82, Gralla16b}.  These are the first results on polar cap properties that include both general relativity and non-zero inclination.  Our findings in the dipole case indicate the presence of pair production, and hence confirm the self-consistency of the dipole model.  The quadrudipole case illustrates the phenomenon of thin annular polar caps, confirming that the shape seen in the aligned case \cite{Gralla16b} persists for non-zero inclination.  For axisymmetric (but possibly inclined) magnetic fields, these properties exhaust the allowed shapes: the polar caps will be either circular or annular.  However, any shape is allowed for intrinsically nonaxisymmetric fields, a case we expect to consider in future work.

The charge-current distribution on the polar cap plays a direct role in several types of pulsar emission.  First, thermal X-ray emission likely arises from heating of the polar cap by pulsar return current \cite{Arons2012}.  Thus, our formulae can be used to model X-ray emission as a function of magnetic field geometry, and hence infer the magnetic geometry from observations (at least in principle).  Our results can also be used to help check the robustness of upcoming neutron star radius measurements by NICER \cite{NICER} to variation in intrinsic magnetic field.  Second, the regions in which electron-positron pair production occurs are determined by bulk current flow \cite{Timokhin13}.  In most models, the pair production is directly related to radio emission and the formation of the pulsar wind.  Given such a model, our results can predict the radio emission signature and pair loading of the pulsar wind as a function of magnetic field geometry.  The gamma-ray emission is likely related to the magnetospheric current sheet \cite{Bai10, Cerutti16}, whose properties are largely insensitive to the magnetic field.  As such, variations between gamma-ray and other types of emission can be a useful probe of magnetic field geometry.

The derivation of our results relies on the spacetime approach to force-free electrodynamics \cite{Gralla14}, making extensive use of differential forms and focusing on invariant properties.  Those readers who are unfamiliar with this approach, or otherwise uninterested in the derivation, may skip directly to Sec.~\ref{sec:Method}, where we state the assumptions and provide a detailed prescription for applying the method.  Finally, in Sec.~\ref{sec:results}, we present results for the dipole and quadrupole pulsars.  Our metric has signature $(-,+,+,+)$ and we use Heaviside-Lorentz units with $G=c=1$.

\section{Method}

We consider the exterior metric of a rotating body, by which we mean a metric with a Killing field $\xi$ that takes the form $\pd_t+\Omega\pd_\varphi$ in cylindrical coordinates far from the star, with $\Omega$ a constant.  We can always choose coordinates such that $\xi$ takes this form everywhere, in which case our assumptions become
\begin{align}\label{eq:HelicalSymmetry}
    \mathcal{L}_\xi g=0,\qquad
    \xi=\pd_t+\Omega\pd_\varphi,
\end{align}
where $\mathcal{L}$ denotes the Lie derivative.  We do not assume at this stage that the spacetime is separately stationary and axisymmetric.  For example, a nonaxisymmetric rotating body (such as a neutron star with a ``mountain'' on it) would still possess the symmetry \eqref{eq:HelicalSymmetry} when gravitational radiation is neglected.  Wherever $\xi$ is timelike, we can define co-rotating observers with four-velocity
\begin{align}
    u^\mu=\xi^\mu/\sqrt{\Upsilon},\qquad
    \Upsilon=-g_{\mu\nu}\xi^\mu\xi^\nu.
\end{align}
We call $\Upsilon$ the redshift factor of the orbit $u^\mu$.  For typical fluid stars, $\xi$ is timelike sufficiently close to the axis of rotation.  The boundary of the timelike region, where $\xi$ becomes null, is called the light cylinder.  It is the radius at which an observer co-rotating with the star would be moving at the speed of light.

Now suppose that the star is magnetized.  If the intrinsic magnetization does not change with time (i.e., the only changes are due to rotation), then the electromagnetic field will share the symmetry of the metric.  Working with the Maxwell two-form $F$, we therefore assume that
\begin{align}\label{eq:HelicalField}
    \mathcal{L}_\xi F=0.
\end{align}
If the electromagnetic field is degenerate ($F\wedge F=0$ or, equivalently, $\bm{E}\cdot\bm{B}=0$), then we may always introduce scalar potentials $\phi_1$ and $\phi_2$ such that \cite{Carter79, Uchida97a, Gralla14}
\begin{align}
    F=\ed\phi_1\wedge\ed\phi_2.
\end{align}
These potentials are the relativistic generalization of Euler potentials.  When the field is magnetically dominated ($F^2>0$ or, equivalently, $\bm{B}^2>\bm{E}^2$), the two-surfaces of constant $\phi_1$ and $\phi_2$ represent worldsheets of magnetic field lines, or ``field sheets'' for short \cite{Gralla14}.  

The symmetry \eqref{eq:HelicalField} implies that $\phi_1$ and $\phi_2$ can be chosen such that \cite{Uchida97b, Gralla14}
\begin{subequations}
\begin{align}
    \phi_1&=\psi_1(r,\theta,\varphi-\Omega t),\\ 
    \phi_2&=\psi_2(r,\theta,\varphi-\Omega t)+\pa{\varphi-\Omega t}\kappa,
\end{align}
\end{subequations}
where $\kappa$ is a constant.  The electric field measured by co-rotating observers is $F\cdot u=F\cdot\xi/\sqrt{\Upsilon}=\kappa\ed\psi_1/\sqrt{\Upsilon}$.  If the star is perfectly conducting, then this must vanish on the surface of the star, thereby fixing $\kappa=0$ (provided $F\neq0$ on the star).  Therefore, we may write
\begin{align}
    F=\ed\psi_1\wedge\ed\psi_2,\qquad
    \psi_i=\psi_i(r,\theta,\varphi-\Omega t).
\end{align}
In particular, the magnetosphere everywhere satisfies
\begin{align}\label{eq:FieldSheet}
    \xi\cdot F=0,
\end{align}
which means that the Killing field is tangent to the field sheets.\footnote{The notation $v\cdot\omega$ indicates the contraction of the vector $v$ into the first index of a differential form $\omega$.  In this case, $\pa{\xi\cdot F}_\nu=F_{\mu\nu}\xi^\mu$.}  In flat spacetime, Eq.~\eqref{eq:FieldSheet} corresponds to the formula $\bm{E}+\bm{V}\times\bm{B}=0$ (see App.~\ref{app:RotatingFlat} for details).  

If the magnetosphere is also force-free ($J\cdot F=0$), then as explained in App.~\ref{app:ConservedQuantity}, Eq.~\eqref{eq:FieldSheet} implies the existence of a quantity $\Lambda$ that is constant on each field sheet.  In terms of the three-current $\bm{J}_c$ and magnetic field $\bm{B}_c$ measured by the co-rotating observers, this conserved quantity may be written as
\begin{align}\label{eq:LambdaCurrent}
    \Lambda(\psi_1,\psi_2)=\sqrt{\Upsilon}J_\parallel,\qquad
    \bm{J}_c=J_\parallel\bm{B}_c.
\end{align}
These formulae hold only inside the light cylinder; outside the light cylinder, $\Lambda$ must be computed in a different manner.

The conserved quantity $\Lambda$ has an amusing cultural history.  The flat spacetime version was discovered by \citet{Mestel73} and subsequently used in \citet{Beskin83}, before being rediscovered by \citet{Uchida98} and re-rediscovered by \citet{Gruzinov05}.  In App.~\ref{app:ConservedQuantity}, we reveal the geometric interpretation and find the full generalization of $\Lambda$: we show that such a conserved quantity always exists in any spacetime $g$ and field configuration $F$ sharing a symmetry $\xi$ that satisfies $\xi\cdot F=0$.

\subsection{Metric}

We now specialize to the case where the stellar surface rotation velocity is much less than the speed of light ($\epsilon\ll1$).  To leading order in rotation, the metric outside of a relativistic star is \cite{Thorne68}
\begin{align}\label{eq:StarMetric}
	ds^2&=-\pa{1-\frac{2M}{r}}\ed t^2+\pa{1-\frac{2M}{r}}^{-1}\ed r^2\\
	&\quad\,+r^2\br{\ed\theta^2+\sin^2{\theta}\pa{\ed\varphi-\Omega_Z\ed t}^2},\quad
	r>R_\star,\nonumber
\end{align}
where the ``frame-drag frequency'' $\Omega_Z$ is
\begin{align}
	\Omega_Z=\frac{2\hat{I}}{r^3}\Omega.
\end{align}
Here, $R_\star$ is the (areal) radius, $M$ is the mass, $\Omega$ is the angular velocity, and $\hat{I}$ is the moment of inertia (defined as the angular momentum over the angular velocity).  The norm of the co-rotation Killing field is
\begin{align}
	\Upsilon=\frac{r-2M}{r}-\br{\pa{\Omega-\Omega_Z}r\sin{\theta}}^2.
\end{align}
The light cylinder is defined to be the locus where $\Upsilon$ vanishes.  We will work with a characteristic light cylinder radius defined by
\begin{align}
	R_L=\frac{1}{\Omega},
\end{align}
which agrees with the actual radius in the slow-rotation limit.

We can characterize the problem by an overall scale and three dimensionless parameters,
\begin{align}\label{eq:StarParameters}
	\epsilon=\frac{R_\star}{R_L},\qquad
	\C=\frac{2M}{R_\star},\qquad
	\I=\frac{\hat{I}}{MR_\star^2},
\end{align}
corresponding to the surface rotation velocity, the stellar compactness, and the dimensionless moment of inertia, respectively.  The compactness has a theoretical upper limit $\C<8/9$ \cite{Buchdahl59}, with realistic neutron star models having $\C\sim1/2$ \cite{Haensel07}.  For clarity of discussion, we will regard the metric \eqref{eq:StarMetric} as exact, although it of course receives higher-order corrections in $\epsilon$.  These corrections do not influence the leading-order calculations performed below.

\subsection{Field strength and expansions}

We wish to solve the equations of force-free electrodynamics perturbatively in $\epsilon$.  Formally, we may imagine having solved the problem at finite $\epsilon$.  This defines a family of solutions
\begin{align}
    F=\ed\psi_1\wedge\ed\psi_2,\qquad
    \psi_i=\psi_i(r,\theta,\varphi-\Omega t;\epsilon).
\end{align}
Perturbation theory consists of Taylor-expanding in $\epsilon$, but we have a choice of what to hold fixed.  We define two expansions,
\begin{subequations}
\begin{align}
    \textrm{near expansion:}&\quad\epsilon\to0\quad\textrm{at fixed }R_\star \\
    \textrm{far expansion:}&\quad\epsilon\to0\quad\textrm{at fixed }R_L,
\end{align}
\end{subequations}
with the other dimensionless parameters $\mathcal{C}$ and $\mathcal{I}$ fixed in both.  We introduce order symbols $\O_\star$ and $\O_L$ to represent the scalings in each limit.  As a trivial example, we have
\begin{subequations}
\begin{align}
    R_\star&=\O_\star(1)=\O_L(\epsilon),\\
    R_L&=\O_\star\pa{\epsilon^{-1}}=\O_L(1). 
\end{align}
\end{subequations}
Other dimensionful parameters scale as
\begin{subequations}
\begin{align}
    M&=\O_\star(1)=\O_L(\epsilon),\\
    \hat{I}&=\O_\star(1)=\O_L\pa{\epsilon^3},\\
    \Omega&=\O_\star(\epsilon)=\O_L(1).
\end{align}
\end{subequations}
(For example, $M=\C R_\star/2=\epsilon\C R_L/2$.)  Since the coordinates are held fixed in both expansions, the scalings of the dimensionful parameters define coordinate regions corresponding to each expansion.  Since the metric and field strength depend on $r/M$, $\hat{I}/r^3$, and $\Omega t$, the regimes of validity are\footnote{Note that $r\gg M$ implies $R\gg R_\star$ because general relativity does not allow stars to become too compact (i.e., we have $R_\star\lesssim M$).  This condition is also implied by the metric \eqref{eq:StarMetric}.}
\begin{align}
    \textrm{near region:}&\quad r\ll R_L,\quad t\ll2\pi/\Omega,\\
    \textrm{far region:}&\quad r\gg R_\star\quad\pa{\textrm{and }\gg M,\hat{I}^{1/3}}.
\end{align}
(The arbitrary choice of working near $t=0$ is inherited from the choice of regarding the Euler potentials as functions of $\varphi-\Omega t$.)  Both expansions are valid in the
\begin{align}
    \textrm{overlap region:}\quad R_\star\ll r\ll R_L,\quad t\ll2\pi/\Omega,
\end{align}
which exists because $\epsilon=R_\star/R_L\ll1$.  This observation allows the large-$r$ behavior of the near expansion to be matched to the small-$r$, small-$t$ behavior of the far expansion.

\subsection{Near zone}

In the near expansion, the geometry reduces to the Schwarzschild metric plus corrections due to rotation,
\begin{align}
	ds^2&=-\pa{1-\frac{2M}{r}}\ed t^2+\pa{1-\frac{2M}{r}}^{-1}\ed r^2\\
	&\quad\,+r^2\br{\ed\theta^2+\sin^2{\theta}\ed\varphi^2}-2\Omega_Z\ed t\ed\varphi
	+\O_\star\pa{\epsilon^2},\nonumber
\end{align}
where we remind the reader that $\Omega_Z=\O_\star(\epsilon)$.  Recalling that $\Omega=\epsilon/R_\star=\O_\star(\epsilon)$, the near expansion of the Euler potentials takes the form
\begin{subequations}
\begin{align}
    \psi_1&=\alpha(r,\theta,\varphi)+\epsilon\br{-\frac{\pd_\varphi\alpha}{R_\star}t+\tilde{\alpha}(r,\theta,\varphi)}
    +\O_\star\pa{\epsilon^2},\\ 
    \psi_2&=\beta(r,\theta,\varphi)+\epsilon\br{-\frac{\pd_\varphi\beta}{R_\star}t+\tilde{\beta}(r,\theta,\varphi)}
    +\O_\star\pa{\epsilon^2},
\end{align}
\end{subequations}
where $\alpha,\beta,\tilde{\alpha},\tilde{\beta}$ are functions of the spatial coordinates only.  Then the field strength is written as
\begin{align}\label{eq:NearField}
    F&=\ed\alpha\wedge\ed\beta+\Omega\ed t\wedge\pa{\pd_\varphi\beta\ed\alpha-\pd_\varphi\alpha\ed\beta}\nonumber\\
    &\quad\,-\Omega t\br{\ed\alpha\wedge\ed\pa{\pd_\phi\beta}+\ed\pa{\pd_\phi\alpha}\wedge\ed\beta}\\
    &\quad\,+\epsilon\pa{\ed\alpha\wedge\ed\tilde{\beta}+\ed\tilde{\alpha}\wedge\ed\beta}
    +\O_\star\pa{\epsilon^2}.\nonumber
\end{align}
where we remind the reader that $\Omega=\epsilon/R_\star=\O_\star(\epsilon)$.  The first term in Eq.~\eqref{eq:NearField} represents the magnetic field of the star when it is not rotating.  We denote it by
\begin{align}
    F^{(0)}=\ed\alpha\wedge\ed\beta.
\end{align}
The remaining terms in \eqref{eq:NearField} are $\O_\star(\epsilon)$.  The second term represents the leading electric field induced by the rotation.  The third term (the second line) contains some time-dependence due to the rotation of the star.  The final term (the third line) represents corrections to the magnetic field of the star.

We will make the basic assumption that the currents are due only to rotation,
\begin{align}\label{eq:CurrentScaling}
    J=\O_\star(\epsilon).
\end{align}
This describes an isolated pulsar, where current flows only because of unipolar induction due to the rotating conductor.  In the aligned case, Eq.~\eqref{eq:CurrentScaling} can be proven from the assumption of asymptotically radial field lines (e.g., Eq.~(12) of Paper I), and it is supported by numerical simulations done in isolation, including those described in App.~\ref{app:Fit}.

Equation~\eqref{eq:CurrentScaling} means that $F^{(0)}$ is a vacuum (no charge or current) Maxwell solution in the Schwarzschild metric.  Since $\alpha$ and $\beta$ are independent of time, this solution is also stationary and purely magnetic.  The stationary, magnetic, vacuum solutions in Schwarzschild spacetime are known in closed form \cite{Anderson70}.  Provided the dipole moment $\mu$ is non-zero, at sufficiently large radius the field always becomes dipolar, and without loss of generality, we may take the dipole moment to instantaneously point along the $x$ axis at an inclination $\iota$ relative to the rotation axis ($z$ axis).  This means that the Euler potentials may always be chosen such that
\begin{subequations}\label{eq:FarLimitOfNear}
\begin{align}
    \alpha&\to\frac{\mu}{r}\sin^2{\theta'}&\textrm{as }r\to\infty,\\
    \beta&\to\varphi'&\textrm{as }r\to\infty,
\end{align}
\end{subequations}
where $\theta'$ and $\varphi'$ are polar coordinates about the dipole axis, related to $\theta$ and $\varphi$ by
\begin{subequations}\label{eq:PrimedCoordinates}
\begin{align}
    \cos{\theta'}&=\cos{\theta}\cos{\iota}-\sin{\theta}\cos{\varphi}\sin{\iota},\\
    \tan{\varphi'}&=\frac{\sin{\theta}\sin{\varphi}}{\sin{\theta}\cos{\varphi}\cos{\iota}+\cos{\theta}\sin{\iota}}.
\end{align}
\end{subequations}
Note that in Eqs.~\eqref{eq:FarLimitOfNear}, the regime $r\to\infty$ does \textit{not} refer to asymptotic infinity because we have already taken the near limit. Instead, $r\to\infty$ refers to the overlap region where we will match to the far zone.

Our choice of a dipole field \eqref{eq:FarLimitOfNear} in the overlap region corresponds to stars whose dipole component of field dominates before the light cylinder is reached, though the dipole can still be sub-dominant at the stellar surface.  For a star with a different moment dominating in the overlap region, Eq.~\eqref{eq:FarLimitOfNear} should be suitably modified to include that moment.  The rest of the analysis presented in this paper (including a new numerical simulation and fit for $\Lambda$) can then be repeated for such a case.

\subsection{Far zone}

In the far expansion, the metric becomes flat,
\begin{align}
    ds^2=-\ed t^2+\ed r^2+r^2\pa{\ed\theta^2+\sin^2{\theta}\ed\varphi^2}
    +\O_L(\epsilon).
\end{align}
The Euler potentials do not simplify, and we simply name the leading piece $\chi_i\equiv\psi_i(r,\theta,\varphi-\Omega t;0)$, so that
\begin{align}
    \psi_i=\chi_i(r,\theta,\varphi-\Omega t)+\O_L(\epsilon).
\end{align} 
Then, the field strength is
\begin{align}
    F&=\ed\chi_1\wedge\ed\chi_2+\O_L(\epsilon).
\end{align}
The small-$r$, small-$t$ behavior of the far expansion must match the large-$r$ behavior of the near expansion.  That is, Eqs.~\eqref{eq:FarLimitOfNear} require the boundary condition
\begin{subequations}\label{eq:NearLimitOfFar}
\begin{align}
    \chi_1&\to\frac{\mu}{r}\sin^2{\theta'}&\textrm{as }r\to0,\\
    \chi_2&\to\varphi'&\textrm{as }r\to0.
\end{align}
\end{subequations}
These equations hold at $t=0$, an arbitrary choice of time at which we chose to align the dipole moment with the $x$ axis [see Eqs.~\eqref{eq:FarLimitOfNear} and \eqref{eq:PrimedCoordinates}].  By the co-rotation symmetry $\xi=\pd_t+\Omega\pd_\varphi$ of the system, similar equations hold at other times, except that $\theta'$ and $\varphi'$ must be aligned with the rotating dipole axis.  Thus, the small-$r$ boundary condition in the far zone is a rotating point dipole.

The large-$r$ boundary condition is that the magnetosphere is isolated.  In practice, it is handled numerically by making the simulation box large enough that edge effects cannot affect the solution before it reaches steady state.

We see that the leading-order field in the far expansion is the force-free magnetosphere of an isolated, rotating, conducting, inclined point dipole in flat spacetime.  This field can be determined numerically by considering a sequence of increasingly smaller rotating conducting stars.  We can then fit for the conserved quantity $\Lambda$ as a function of $\chi_1$ and $\chi_2$.  In the numerical simulation, this is most conveniently done near or on the star, which by definition is in the overlap region where $\chi_1=\alpha$ and $\chi_2=\beta$ of the matched asymptotic expansion [see Eqs.~\eqref{eq:FarLimitOfNear} and \eqref{eq:NearLimitOfFar}].

In finding a suitable fit, our goal is to capture the qualitative features while maintaining simplicity.  We find that an excellent fit (see App.~\ref{app:Fit}) is given by
\begin{align}\label{eq:Fit}
    \Lambda(\alpha,\beta)=\mp2\Omega\Big\{ &J_0\pa{2\arcsin{\sqrt{\alpha/\alpha_o}}}\cos{\iota}\nonumber\\ 
    &-J_1\pa{2\arcsin{\sqrt{\alpha/\alpha_o}}}\cos{\beta}\sin{\iota}\Big\},\nonumber\\
    &\qquad\alpha<\alpha_o, 
\end{align}
where $J_0$ and $J_1$ are Bessel functions of the first kind and the ``last open field line'' $\alpha_o$ is given by
\begin{align}\label{eq:LastOpenFieldLine}
    \alpha_o=\sqrt{\frac{3}{2}}\mu\Omega\pa{1+\frac{1}{5}\sin^2{\iota}}.
\end{align}
The upper/lower sign in Eq.~\eqref{eq:Fit} corresponds to the northern/southern flow.

Our motivations for this fit are the following.  In the limit $\epsilon\to0$, the polar cap occupies a vanishingly small portion of the sphere, and may therefore be approximated as a disk.  The natural fitting functions are then the Bessel harmonics $J_n(\rho)e^{\pm in\gamma}$ (for $n\geq 0$ with $(\rho,\gamma)$ polar coordinates in the disk).  These harmonics form representations of the Euclidean group $\mathsf{E}_2$ and are the disk analogs of spherical harmonics.  The angular coordinate $\gamma$ must equal $\phi'=\beta$ by axisymmetry.  The radial coordinate $\rho$ should be proportional to $\theta'$, but the proportionality constant is free since we have not yet fixed the size of the disk.  Noting that $\alpha\propto\sin^2{\theta'}$ because the field is dipolar in this region, we have $\theta'=\arcsin{\sqrt{\alpha/\alpha_o}}$ for some constant $\alpha_o$.  We find that $\rho=2\theta'$ works well, resulting in the argument $2\arcsin\sqrt{\alpha/\alpha_o}$ for the Bessel functions.  The $\arcsin$ function is undefined for $\alpha>\alpha_o$, so $\alpha_o$ naturally delineates the boundary of the polar cap.  We fit the simple functional form \eqref{eq:LastOpenFieldLine} to the size of the polar cap.  Remarkably, an excellent fit for the current $\Lambda$ is now obtained by simply taking a linear combination of the Bessel harmonics with prefactors $\cos{\iota}$ and $\sin{\iota}$, as shown in Eq.~\eqref{eq:Fit}.

Interestingly, an analogous expression involving Bessel functions provides a qualitative description \cite{Gralla15} of the force-free jets produced by a spinning black hole embedded in a misaligned magnetic field \cite{Palenzuela10}.

The simple fit \eqref{eq:Fit} is adequate for the applications we have in mind, and hence we use it for the remainder of the paper.  However, we emphasize that improvements to the simulation and fitting could surely produce a more accurate expression for $\Lambda(\alpha,\beta)$.  It is straightforward to repeat the analysis of the rest of the paper for any such expression.

\subsection{Near-zone charge-current}

We may now determine the near-zone four-current to leading order $\O_\star(\epsilon)$.  We can obtain three components from the conservation law $\bm{J}=\pa{\Lambda/\sqrt{\Upsilon}}\bm{B}$ [see Eq.~\eqref{eq:LambdaCurrent}].  Since $\Lambda=\O_\star(\epsilon)$ and $\bm{J}=\O_\star(\epsilon)$, we need only the $\O_\star(1)$ piece of $\bm{B}$ (i.e., $F^{(0)}=\ed\alpha\wedge\ed\beta$) to determine the leading, $\O_\star(\epsilon)$ current.  A straightforward calculation yields
\begin{subequations}\label{eq:Current}
\begin{align}
    J^r&=\frac{\Lambda(\alpha,\beta)}{r^2\sin{\theta}}F_{\theta\phi}^{(0)}
    +\O_\star\pa{\epsilon^2},\\
    J^\theta&=\frac{\Lambda(\alpha,\beta)}{r^2\sin{\theta}}F_{\phi r}^{(0)}
    +\O_\star\pa{\epsilon^2},\\
    J^\phi&=\frac{\Lambda(\alpha,\beta)}{r^2\sin{\theta}}F_{r\theta}^{(0)} + \O_\star\pa{\epsilon^2}.
\end{align}
\end{subequations}
The fourth component can be obtained by taking the divergence $J^\nu=\nabla_\mu F^{\mu\nu}$ directly from Eq.~\eqref{eq:NearField},
\begin{align}\label{eq:ChargeDensity}
    J^t&=\frac{\Omega-\Omega_Z}{r\pa{r-2M}}
    \bigg\{\pd_\theta\alpha\pd_\theta\pd_\phi\beta
        -\pd_\theta\beta\pd_\theta\pd_\phi\alpha\\
    &\quad\,+r\pa{r-2M}\pa{\pd_r\alpha\pd_r\pd_\phi\beta
        -\pd_r\beta\pd_r\pd_\phi\alpha}\nonumber\\
    &\quad\,-\pd_\phi\alpha\br{\pa{1-\frac{2M}{r}}\pd_r\pa{r^2\pd_r\beta}+\frac{\pd_\theta\pa{\sin{\theta}\pd_\theta\beta}}{\sin{\theta}}}\nonumber\\
    &\quad\,+\pd_\phi\beta\br{\pa{1-\frac{2M}{r}}\pd_r\pa{r^2\pd_r\alpha}+\frac{\pd_\theta\pa{\sin{\theta}\pd_\theta\alpha}}{\sin{\theta}}}
    \bigg\}\nonumber.
\end{align}
Notice that the result involves only $\alpha$ and $\beta$; the higher-order corrections $\tilde{\alpha}$ and $\tilde{\beta}$ do not appear.\footnote{In flat spacetime, this corresponds to the statement that we may compute $\rho_e=\bm{\nabla}\cdot\bm{E}$ from knowledge of $\bm{B}$ only, because $\bm{E}=-\bm{V}\times\bm{B}$ with $\bm{V}=\rho\hat{\bm{\phi}}$.}  This is essential to the method, since the higher-order corrections are unknown.  Similar direct computations of the spatial components would result in expressions featuring $\tilde{\alpha}$ and $\tilde{\beta}$.  These expressions could be set equal to the current found from the conservation of $\Lambda$ [Eq.~\eqref{eq:Current}] to provide equations for the corrections $\tilde{\alpha}$ and $\tilde{\beta}$.

According to Eqs.~\eqref{eq:Current} and \eqref{eq:ChargeDensity}, the charge-current is only non-zero where the conserved quantity $\Lambda(\alpha,\beta)$ is non-zero.  There is a subtlety, however, in that $\Lambda$ is a conserved quantity along field sheets (equivalently, field lines), but $\alpha$ and $\beta$ may not uniquely label the field lines.  The charge-current \eqref{eq:Fit}-\eqref{eq:LastOpenFieldLine} is determined in the overlap region, and hence charge and current flow only on the field lines that reach the overlap region (which are also the field lines that open up to asymptotic infinity).  Thus, in case there are multiple field lines with $\alpha<\alpha_o$ in the near region, one is to use Eqs.~\eqref{eq:Current} and \eqref{eq:ChargeDensity} \textit{only} for the field lines that reach the overlap region where the labeling becomes unique.  On all other field lines, the current flow is zero.  This ensures that there are only two polar caps; see Paper I for further discussion of the axisymmetric case.

\section{Statement of method}
\label{sec:Method}

\begin{figure*}
    \includegraphics[width=\textwidth]{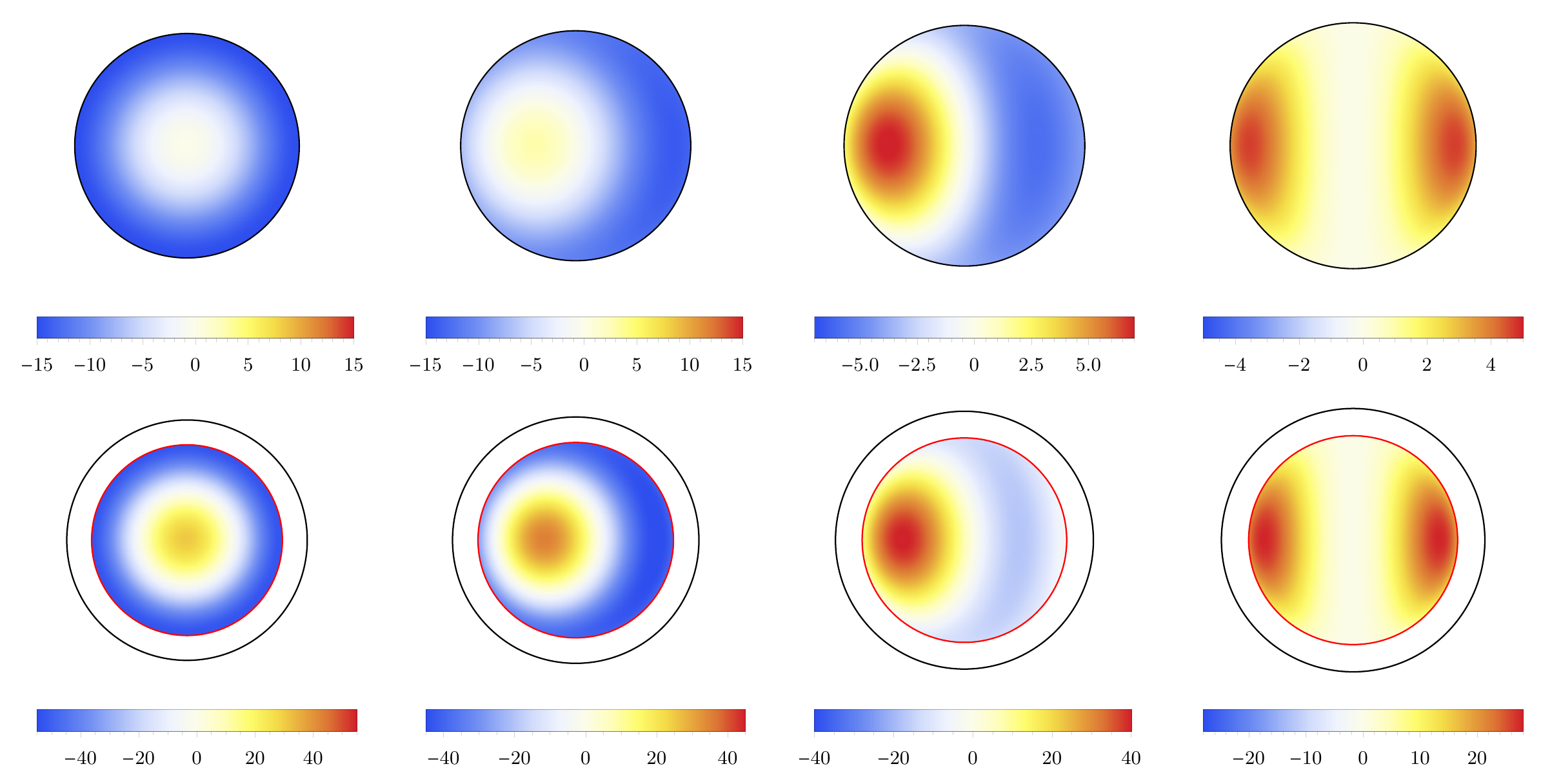}
    \caption{Polar cap structure for the dipole pulsar.  The polar caps have a circular shape near the magnetic poles, occupying a physical area scaling as $\epsilon^2$.  We show $J^2R_\star^2/\pa{\epsilon B_1}^2$, the dimensionless charge-current norm with the rotation rate scaled out.  (Here, as in Paper I, $B_1$ is one-half the value of the magnetic field on the pole.)  From left to right, we display inclinations $\iota=\cu{0^\circ,30^\circ,60^\circ,90^\circ}$.  In the bottom row, we use realistic parameters $\C=1/2$ and $\I=2/5$, while in the top row, we set $\C=\I=0$, which corresponds to neglecting general relativity.  The polar cap shrinks with increasing stellar compactness $\C$, as illustrated by the black circles showing the $\C=0$ polar cap size for each respective inclination.}
    \label{fig:DipolePolarCaps}
\end{figure*}

\begin{figure}
    \includegraphics[width=.4\textwidth]{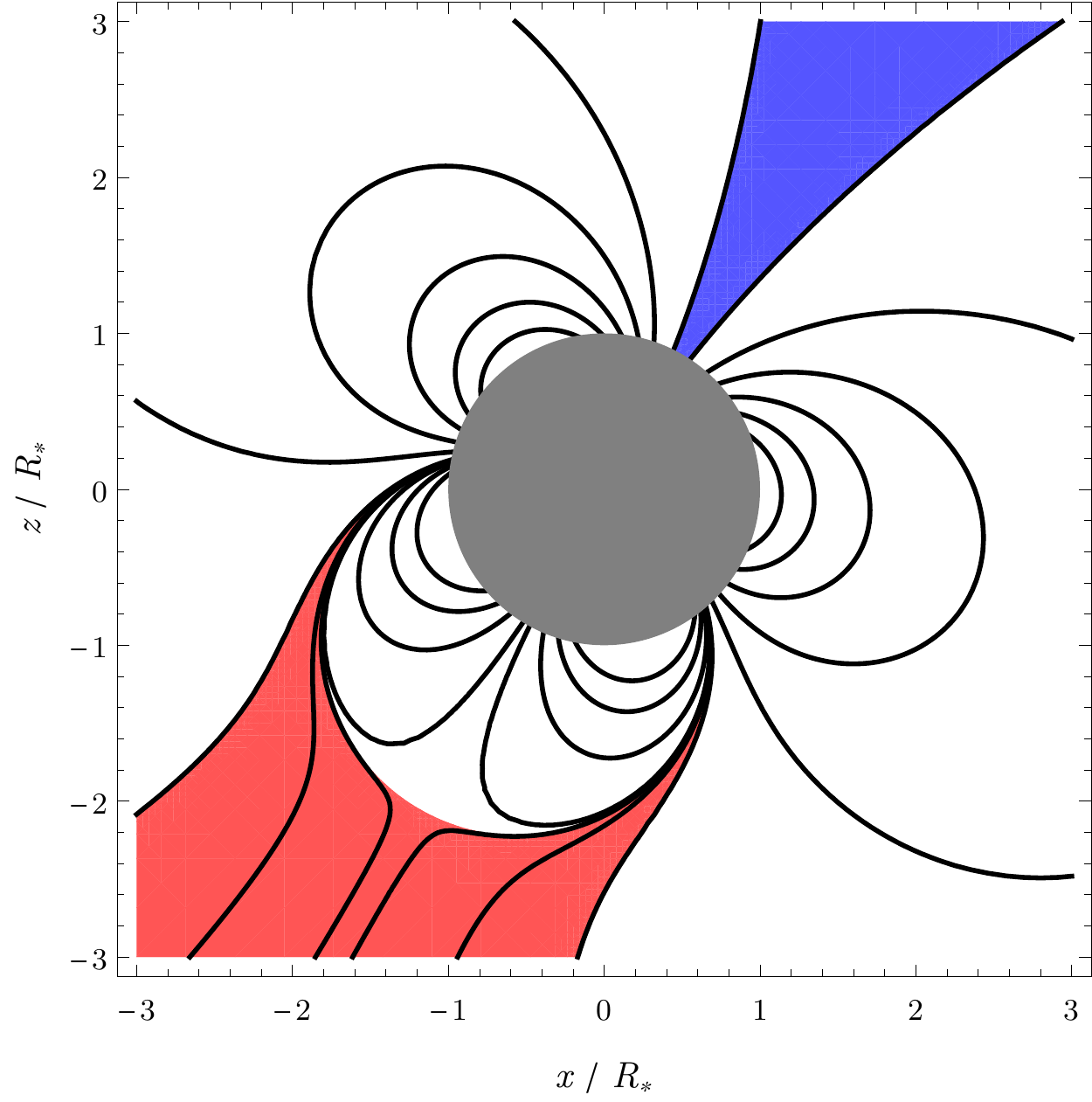}
    \caption{Poloidal field lines (level sets of $\alpha$) for the quadrudipole pulsar with the quadrupole-to-dipole ratio $q=3$ and inclination $\iota=30^\circ$.  We shade in the regions of current flow for $\epsilon=1/50$ (as in Fig.~4 of Paper I).  The intersections of these flows with the stellar surface define the polar caps.  We see that the northern (blue) cap is circular, while the southern (red) cap is annular.}
    \label{fig:Quadrudipole}
\end{figure}

We now summarize the assumptions and describe the method we have derived.  We have assumed that:
\begin{enumerate}
\item The star is spherical (mass $M$ and areal radius $R_\star$) and rigidly rotating with angular velocity $\bm{\Omega}=\Omega\hat{\bm{z}}$ such that $\Omega R_\star\ll c$.
\item The stellar surface is perfectly conducting and the magnetosphere is force-free.  The star is isolated.
\item The star is magnetized in such a way that when it is not rotating, the magnetic field is dipolar at large distances.  Furthermore, the angular velocity is small enough that the dipole component dominates before the light cylinder radius $r\sim R_L=c/\Omega$ is reached.
\end{enumerate}
Under these assumptions, we provide analytic formulae for the near-field charge and current associated with a given choice of stellar parameters \eqref{eq:StarParameters} and magnetization.  The magnetization is described by a static, vacuum, asymptotically dipolar magnetic field solution in the  Schwarzschild spacetime.  This solution must be expressed in terms of Euler potentials $\alpha(t,r,\phi)$ and $\beta(t,r,\phi)$ in Schwarzschild coordinates.  (That is, the Schwarzschild coordinate components of $F$ are given by $F_{\mu\nu}=\pd_\mu\alpha\pd_\nu\beta-\pd_\nu\alpha\pd_\mu\beta$.)  For axisymmetric fields (not necessarily aligned with the rotation axis), one can accomplish this by picking an axisymmetric flux function $\psi(r,\theta)$ (e.g., App.~B2 of Ref.~\cite{Gralla16a}) and setting
\begin{align}\label{eq:AlphaBeta}
    \alpha=\psi(r,\theta'),\qquad
    \beta=\varphi',
\end{align}
where $(\theta',\varphi')$ denote spherical coordinates rotated to the desired inclination angle $\iota$ [see Eqs.~\eqref{eq:PrimedCoordinates}].

The Euler potentials $\alpha$ and $\beta$ label magnetic field lines.  The labeling is not necessarily one-to-one, but it becomes so at sufficiently large $r$ as the field becomes dipolar.  Current flows on the field lines that have $\alpha<\alpha_o$ [see Eq.~\eqref{eq:LastOpenFieldLine}] \textit{and} enter the region where the labeling is unique.\footnote{Near the star, there can be additional field lines labeled by $\alpha<\alpha_o$; these field lines do not connect to the far zone, and hence no current flows on them.}  The portion of space occupied by these field lines has non-zero charge and current given by\footnote{Here, we present orthonormal-frame components of static Schwarzschild observers, using the notation and definitions of Ch.~II of Ref.~\cite{Thorne86}.  This agrees with the measurements of co-rotating observers to leading order in the rotation $\Omega R_\star$.}
\begin{align}
    \rho_e&=J^{\hat{t}},\qquad
    \bm{J}=J^{\hat{r}}\hat{\bm{r}}+J^{\hat{\theta}}\hat{\bm{\theta}}+J^{\hat{\phi}}\hat{\bm{\phi}},
\end{align}
where 
\begin{subequations}
\begin{align}
    J^{\hat{t}}&=\sqrt{1-\frac{2M}{r}}J^t,\\
    J^{\hat{r}}&=\frac{\Lambda(\alpha,\beta)}{\sqrt{r\pa{r-2M}}\pa{r\sin{\theta}}}
    \pa{\pd_\theta\alpha\pd_\phi\beta-\pd_\phi\alpha\pd_\theta\beta},\\
    J^{\hat{\theta}}&=\frac{\Lambda(\alpha,\beta)}{r\sin{\theta}}
    \pa{\pd_\phi\alpha\pd_r\beta-\pd_r\alpha\pd_\phi\beta},\\
    J^{\hat{\phi}}&=\frac{\Lambda(\alpha,\beta)}{r}
    \pa{\pd_r\alpha\pd_\theta\beta-\pd_\theta\alpha\pd_r\beta},
\end{align}
\end{subequations}
with $J^t$ given in Eq.~\eqref{eq:ChargeDensity} and $\Lambda$ given in Eq.~\eqref{eq:Fit}.  These formulae express the charge and current at leading order in $\epsilon$ near the star ($r\ll R_L$).  When evaluated on the star, they provide the polar cap structure.

\section{Results: dipole and quadrudipole}\label{sec:results}

\begin{figure*}
    \includegraphics[width=.246\textwidth]{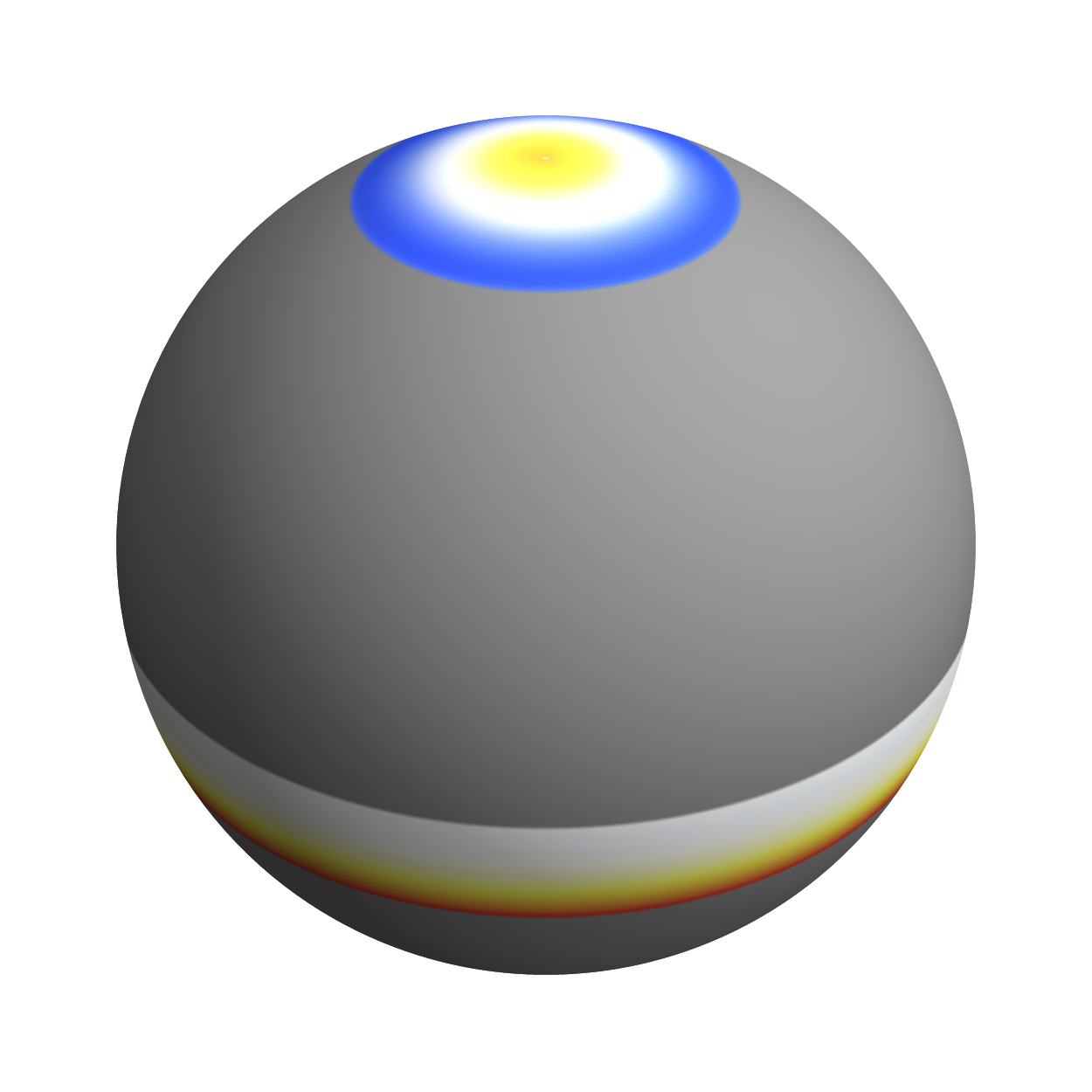}
    \includegraphics[width=.246\textwidth]{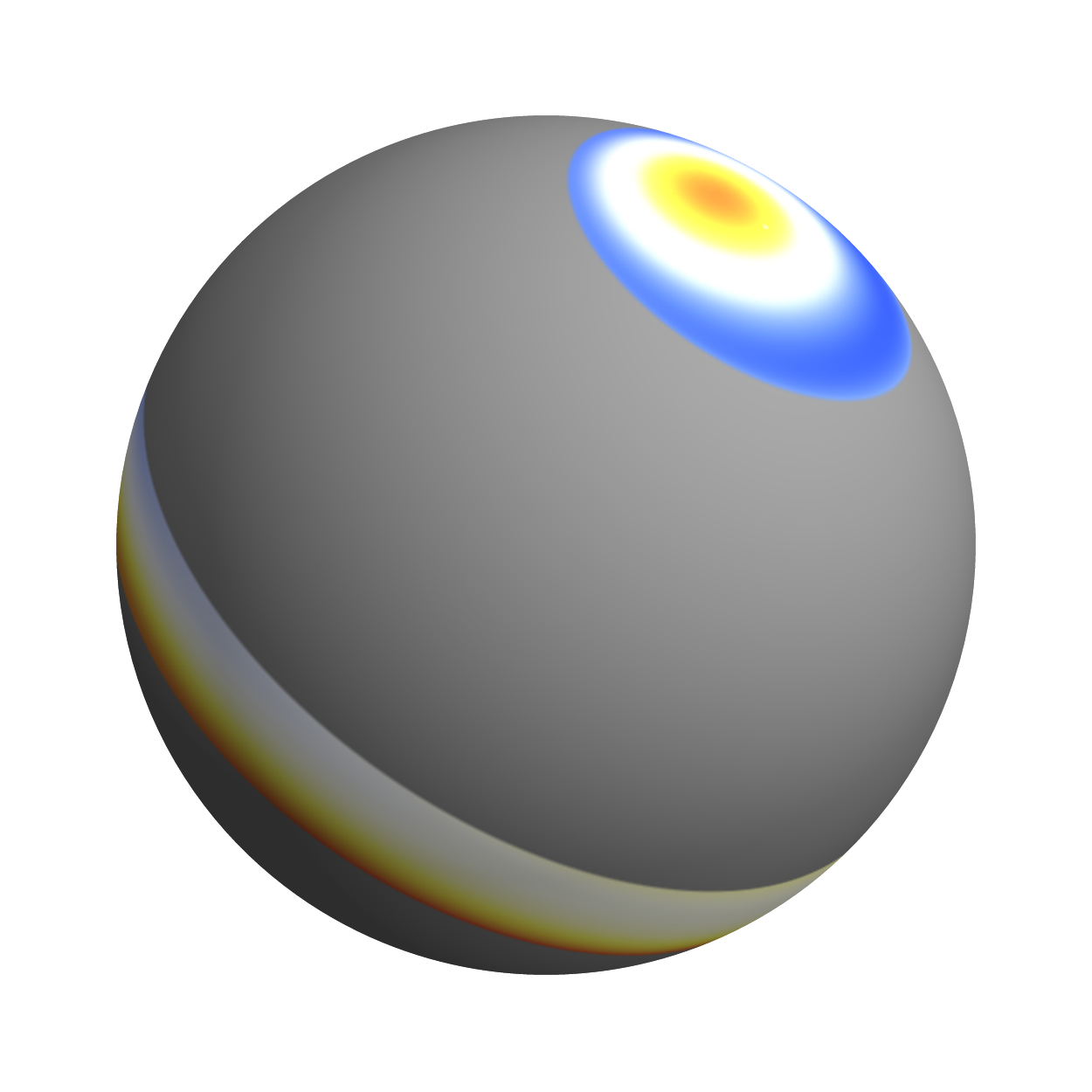}
    \includegraphics[width=.246\textwidth]{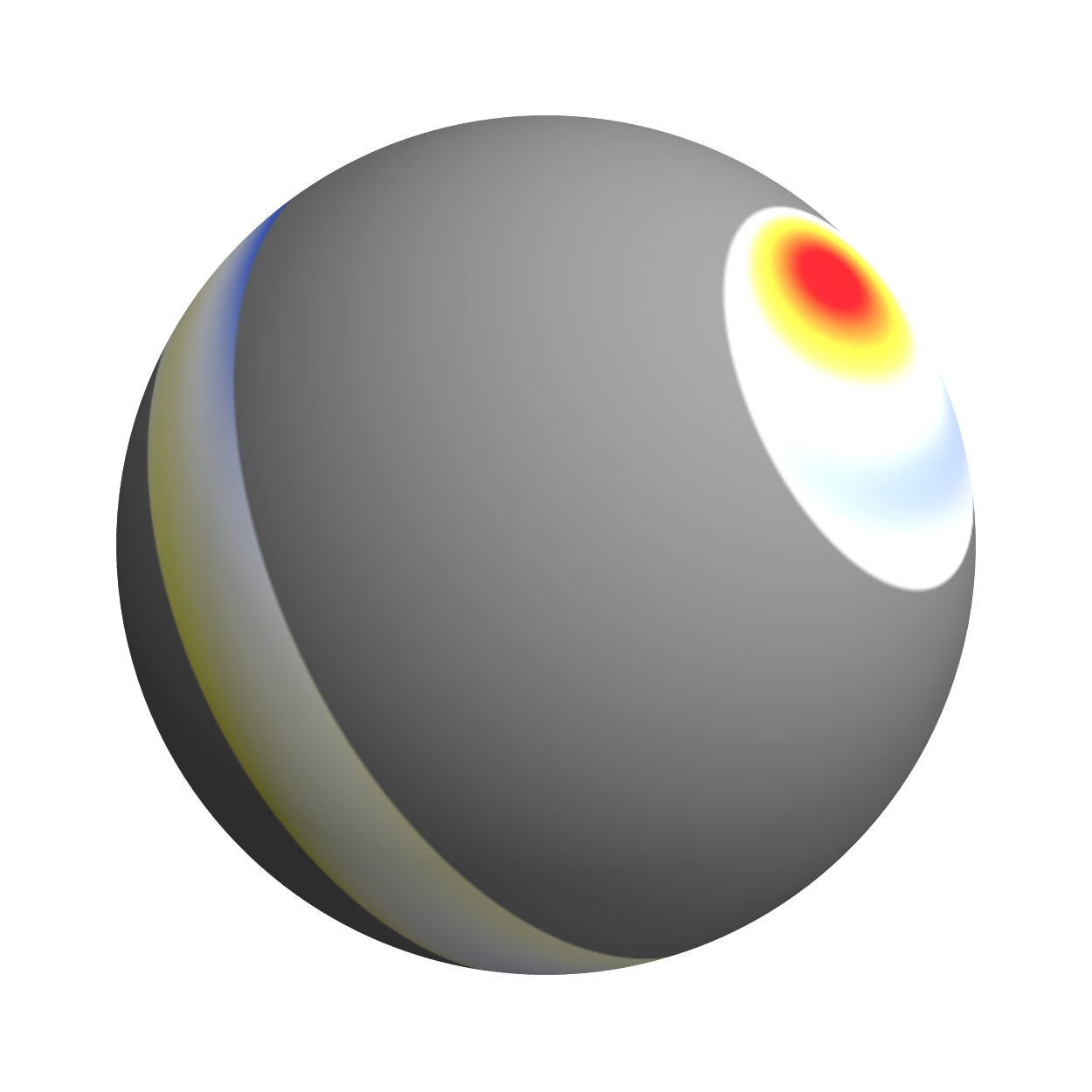}
    \includegraphics[width=.246\textwidth]{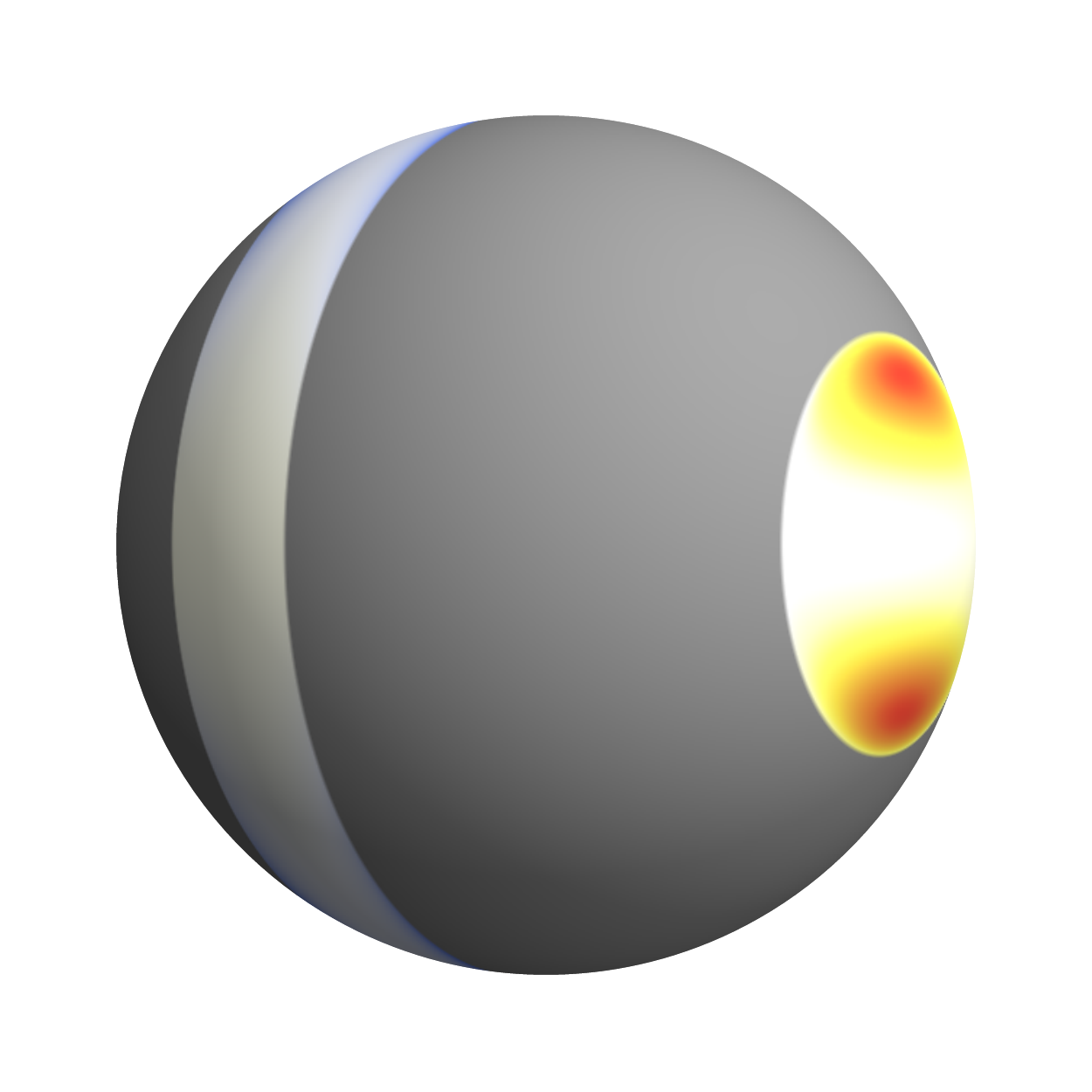}

    \vspace{-10pt}
    \includegraphics[width=\textwidth]{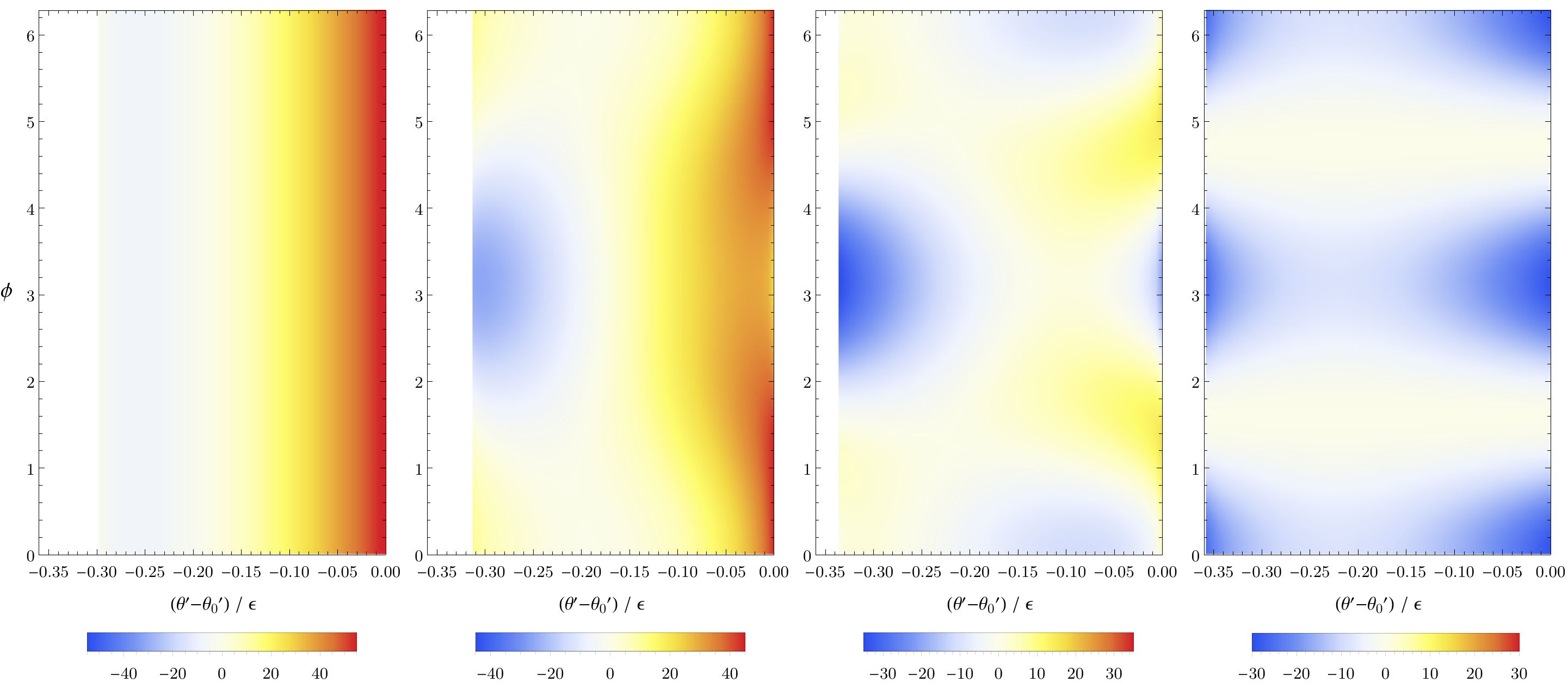}
    \caption{Polar cap structure for the quadrudipole pulsar with the quadrudipole-to-dipole ratio $q=3$, moment of inertia $\I=2/5$, and compactness $\C=1/2$.  From left to right, we display inclinations $\iota=\cu{0^\circ,30^\circ,60^\circ,90^\circ}$.  The color map corresponds to the charge-current norm $J^2R_\star^2/\pa{\epsilon B_1}^2$, as described in Fig.~\ref{fig:DipolePolarCaps}.  In the top row, we show the distribution on the sphere (gray means zero charge and current), with the polar caps made artificially large (by scaling with the relative area kept invariant) for illustration purposes.  (The area of each cap scales as $\epsilon$.  Even for the fastest rotating pulsars, with $\epsilon\sim1/5$, the southern cap would be only $3^\circ$ wide.)  In the bottom row, we show a zoomed-in view of the annular southern polar cap with $\epsilon$ scaled out.  The southern edge of the cap lies at an angle $\theta'=\theta_0'\approx109^\circ$.  \href{https://youtu.be/M_ruTbM8YNo}{3D animations are available here.}}
    \label{fig:SouthernStrips}
\end{figure*}

We now present results for two specific choices of stellar magnetization.  First, we consider the canonical case of a pure dipole.  The flux function $\psi$ is given (e.g.) in Eq.~(40) of Paper I.  Following Eq.~\eqref{eq:AlphaBeta}, the Euler potentials are given by
\begin{subequations}
\begin{align}
    \alpha&=-\frac{3}{2r}\br{\frac{3+4\log{f}-4f+f^2}{\pa{1-f}^3}}\mu\sin^2{\theta'},\\
    \beta&=\varphi',\qquad\qquad\qquad\qquad\quad
    f=1-\frac{2M}{r}.
\end{align}
\end{subequations}
(Although it is not apparent, this solution does approach $\mu\sin^2{\theta'}/r$ at large $r$, as required.)  In Fig.~\ref{fig:DipolePolarCaps}, we plot the norm of the charge-current, $J^2=J_\mu J^\mu=\bm{J}^2-\rho_e^2$, for a variety of parameters.  The main result is that for realistic compactness, there are regions of spacelike current ($J^2>0$) at all inclinations.  Since spacelike current ensures the pair production necessary to sustain a high-multiplicity magnetosphere  \citep{Timokhin13, Chen14, Philippov15a, Philippov15b}, our results support the basic consistency of the dipole force-free model.  We also show flat spacetime (zero-compactness) results to compare with previous work and illustrate the role of general relativity.  These are the first results for the polar cap structure of the general relativistic force-free inclined dipole pulsar.

As a more complicated example, we consider a quadrudipole field (a superposition of dipole and quadrupole fields).  We choose the moment ratio $q=3$ in the notation of Sec.~IV of Paper I.  The flux function $\psi(\theta)$ (and hence the Euler potential $\alpha=\psi(\theta')$) is given by Eq.~(50) of Paper I.  Instead of reproducing the formula, we include a plot of the level sets of $\alpha$ (i.e., the poloidal field lines) in Fig.~\ref{fig:Quadrudipole}.

The current norm $J^2$ is plotted in Fig.~\ref{fig:SouthernStrips} for a variety of inclinations.  Notice that the southern polar cap takes a thin annular shape, a typical feature of non-dipolar fields.  We refer the reader to Paper I for a discussion of the potential importance of this feature, including the possibility that it accounts for the modified beam characteristics of millisecond pulsars.

We now discuss an interesting feature of the orthogonal case $\iota=90^\circ$.  Notice that when the polar cap is circular (dipole or northern cap of quadrudipole), the current is entirely spacelike, while for the annular cap there are both timelike and spacelike currents.  This can be understood from the fact that the charge density scales like $\rho_e\sim\pa{\bm{\Omega}-\bm{\Omega}_Z}\cdot\bm{B}$.  In the orthogonal case, the circular polar caps lie in a region where the rotation and magnetic field become orthogonal, making the charge density anomalously small.  Thus, the four-current is entirely spacelike.  On the other hand, the annular polar cap is shifted away from this region, and both kinds of currents are present.

Finally, we discuss the presence of volume return current.  By ``return current'' we mean a region near the star satisfying $\mathbf{J} \cdot \mathbf{n}/\rho_e<0$, where $\mathbf{n}$ is the co-rotating normal vector to the star and the charge and current are those measured by co-rotating observers.  This corresponds to inward flow of plasma when only a single sign of charge is present.  Noting that $\Lambda\sim\bm{J}\cdot\bm{B}$, the regions of volume return current are those with $\pm\Lambda/\rho_e<0$, where $+$ and $-$ correspond to the northern ($\mathbf{B}\cdot \mathbf{n}>0$) and southern ($\mathbf{B}\cdot \mathbf{n}<0$) caps, respectively.

We describe the qualitative features rather than presenting plots.  For non-orthogonal inclinations, the charge density does not change sign on the polar caps.  We find that there is a small region of volume return current at the outer edge (the edge further from the magnetic pole) in all the non-orthogonal cases we consider.  In the orthogonal case, the charge density changes sign, but we find that there is no volume return current for both the dipolar and quadrudipolar magnetic geometries.

\section*{Acknowledgements}

We thank Feryal \"Ozel, Dimitrios Psaltis, Anatoly Spitkovsky, and Alexander Tchekhovskoy for helpful conversations.  This work was supported by NSF grants 1205550 to Harvard University and 1506027 to the University of Arizona, by the NASA Earth and Space Science Fellowship Program (grant NNX15AT50H to A.P.) and by the Porter Ogden Jacobus Fellowship, awarded to A.P. by the graduate school of Princeton University. The simulations presented in this paper used computational resources supported by the PICSciE-OIT High Performance Computing Center and Visualization Laboratory, and by the NASA/Ames HEC Program (SMD-16-6663, SMD-16-7816).

\appendix

\section{Conserved quantity}
\label{app:ConservedQuantity}

\begin{figure*}
    \includegraphics[width=\textwidth]{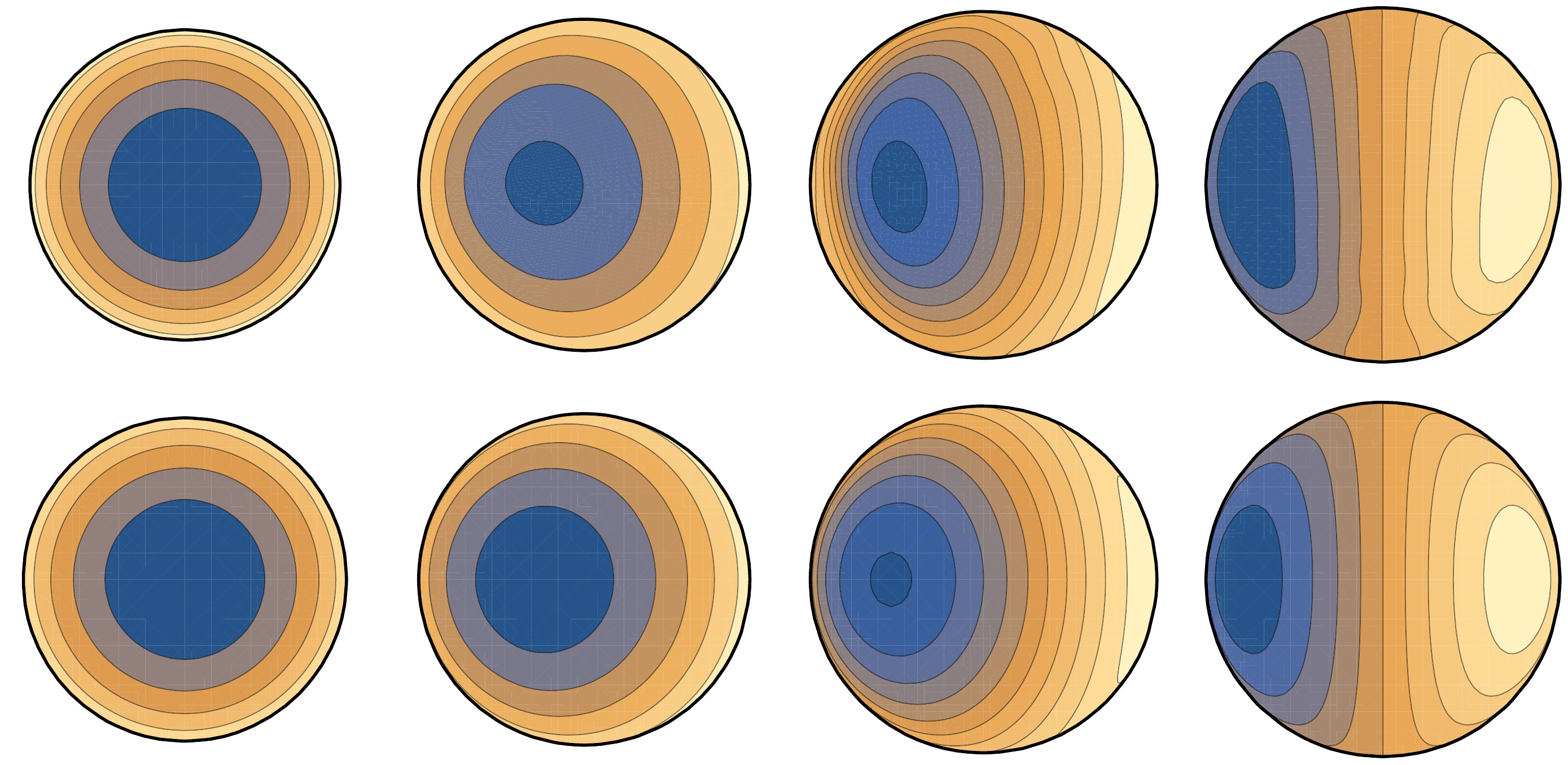}
    
    \includegraphics[width=\textwidth]{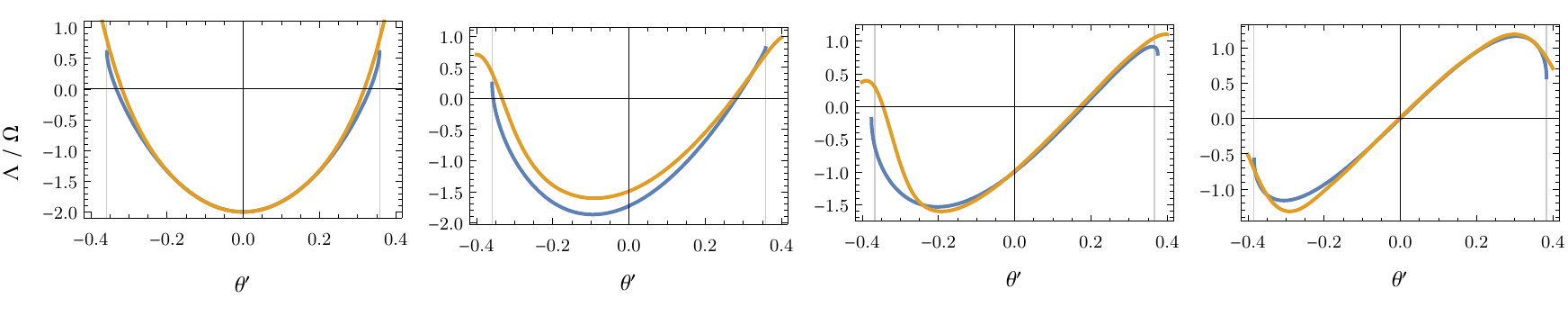}
    
    \vspace{-10pt}
    \caption{Comparison of numerical results for $\Lambda$ with the analytic fit \eqref{eq:Fit}.  From left to right, we display inclinations $\iota=\cu{0^\circ,30^\circ,60^\circ,90^\circ}$.  We show contours for the numerical results (top row) and analytical fit (middle row) for the northern polar cap.  The bottom row shows a cross section, with numerical results in blue and the analytic fit in orange.  The vertical lines indicate the automatic cutoff imposed by the $\arcsin$ function in the analytic fit, delineating the size \eqref{eq:LastOpenFieldLine} of the polar cap.  In these plots the ``numerical results'' for the aligned case are actually the high-precision fit presented in Paper I.}
    \label{fig:Fit}
\end{figure*}

Force-free electrodynamics in an arbitrary curved spacetime may be expressed in terms of three scalar potentials $(\phi_1,\phi_2,\lambda)$ as (see footnote 6 of Ref.~\cite{Gralla14})
\begin{align}
    \ed\star F=\ed\lambda\wedge F,\qquad
    F=\ed\phi_1\wedge\ed\phi_2.
\end{align}
Here, $\phi_1$ and $\phi_2$ are relativistic Euler potentials.  There is no standard name for $\lambda$, but it acts as a stream function for the charge-current as measured in units of magnetic field strength.  In particular, an observer with four-velocity $u^\mu$ measures current density $\bm{J}$ and magnetic field $\bm{B}$ related by
\begin{align}
    \bm{J}=J_\parallel\bm{B},\qquad
    J_\parallel=-u\cdot\ed\lambda.
\end{align}
From studies of stationary force-free magnetic fields, we are used to this field-aligned current $J_\parallel$ being conserved along field lines.  To generalize this statement, we suppose that the metric and field configuration have a symmetry tangent to the field sheets,
\begin{align}\label{eq:SymmetryAssumptions}
    \mathcal{L}_\xi g=0,\qquad
    \mathcal{L}_\xi F=0,\qquad
    \xi\cdot F=0.
\end{align}
The current three-form $J=\ed\star F$ will also respect the symmetry, so
\begin{subequations}
\begin{align}
    0&=\mathcal{L}_\xi J=\xi\cdot\ed J+\ed\pa{\xi\cdot J}=\ed\pa{\xi\cdot J}\\
    &=\ed\br{\xi\cdot\pa{\ed\lambda\wedge F}}=\ed\br{\pa{\xi\cdot\ed\lambda}\wedge F}\\
    &=\ed\pa{\xi\cdot\ed\lambda}\wedge\ed\phi_1\wedge\ed\phi_2.
\end{align}
\end{subequations}
This means that $\xi\cdot\ed\lambda$ is a function only of $\phi_1$ and $\phi_2$,
\begin{align}
    \xi\cdot\ed\lambda=-\Lambda(\phi_1,\phi_2).
\end{align}
If $\xi$ is timelike, then we may define an associated observer $\O$ with four-velocity $u^\mu$ and redshift factor $1/\sqrt{\Upsilon}$,
\begin{align}
    u^\mu=\xi^\mu/\sqrt{\Upsilon},\qquad
    \Upsilon=-g_{\mu\nu}\xi^\mu \xi^\nu. 
\end{align}
It is then straightforward to show that the magnetic field $\bm{B}_\O$ and three-current $\bm{J}_\O$ seen by these observers satisfy
\begin{align}\label{eq:ObserverFieldCurrent}
    \bm{B}_\O=\pa{\Lambda/\sqrt{\Upsilon}}\bm{J}_\O.
\end{align}
Thus, in regions where the Killing field $\xi$ is timelike, we may interpret the conserved quantity $\Lambda$ as a redshifted field-aligned current.  (Note that there is no component of field perpendicular to the current, as guaranteed by force-free electrodynamics.)  Note also that by Eq.~\eqref{eq:SymmetryAssumptions}, we may always choose the potentials to satisfy \cite{Uchida97b, Gralla14}
\begin{align}\label{eq:GaugeChoice}
    \xi\cdot\ed\phi_1=0,\qquad
    \xi\cdot\ed\phi_2=0.
\end{align}
We next compare to previous work, showing how this conserved quantity generalizes those previously known in various different contexts.

\subsection{Force-free magnetic fields}

If spacetime is flat and $\xi=\pd_t$ is the time-translation Killing field, then our assumptions \eqref{eq:SymmetryAssumptions} correspond to
\begin{align}
    \pd_t\bm{B}=0,\qquad
    \bm{E}=0,
\end{align}
a setup generally known as a force-free magnetic field.  In this case, $\Upsilon=1$ identically, so the conserved quantity $\Lambda$ is just the field-aligned current.  By Eq.~\eqref{eq:GaugeChoice}, we may always choose $\phi_1$ and $\phi_2$ to be independent of $t$, making them the usual Euler potentials, which label the field lines as $\bm{B}=\bm{\nabla}\phi_1\times\bm{\nabla}\phi_2$.  Thus, we recover the standard result that $J_\parallel$ is constant on field lines.

\subsection{Rotating configuration in flat spacetime}
\label{app:RotatingFlat}

If $\xi=\pd_t+\Omega\pd_\phi$ in flat spacetime, then the assumptions \eqref{eq:SymmetryAssumptions} correspond to 
\begin{align}\label{eq:RotatingFlat}
    \bm{B}=\bm{B}(\rho,z,\phi-\Omega t),\qquad
    \bm{E}+\bm{V}\times\bm{B}=0,
\end{align}
where $\bm{V}=\omega\rho\hat{\bm{\phi}}$. Inside the light cylinder, $\rho<\Omega^{-1}$, we may define co-rotating observers with a four-velocity and redshift
\begin{align}
    \sqrt{\Upsilon}u^\mu_{\rm corot}=\pd_t+\Omega\pd_\phi,\qquad
    \Upsilon=1-\pa{\rho\Omega}^2.
\end{align}
The current density and magnetic field according to these observers are
\begin{align}
    \sqrt{\Upsilon}\bm{J}_{\rm corot}=\bm{J}-\rho_e\Omega\rho\hat{\bm{\phi}},\qquad
    \bm{B}_{\rm corot}=\bm{B},
\end{align}
where $\bm{J}$ and $\rho_e$ are the current and charge densities in the fixed frame.  Thus, by Eq.~\eqref{eq:ObserverFieldCurrent}, we may write
\begin{align}\label{eq:Uchida}
    \bm{J}-\rho_e\Omega\rho\hat{\bm{\phi}}=\Lambda\bm{B},
\end{align}
the form given by \citet{Uchida98}.  The left side of this equation is sometimes called the current in a rotating frame.  We will avoid this terminology since we reserve ``frame'' for the physical measurements of some class of observers.  An alternative form is given by \citet{Gruzinov06},
\begin{align}\label{eq:Gruzinov}
    \bm{\nabla}\times\br{\bm{B}+\bm{V}\times\pa{\bm{V}\times\bm{B}}}=\Lambda\bm{B}.
\end{align}
Equations~\eqref{eq:Uchida} and \eqref{eq:Gruzinov} are equivalent under the assumptions \eqref{eq:RotatingFlat} (and the force-free equations).

\subsection{Axisymmetric solutions}

In a general axisymmetric (circular) spacetime, one may characterize axisymmetric solutions with non-zero poloidal field by the flux function $\psi(r,\theta)$, the polar current $I(\psi)$, and field line velocity $\Omega(\psi)$.  We adopt the conventions of Paper I, where the flux function is the magnetic flux through a loop of revolution divided by $2\pi$, and the polar current is minus the current through the loop of revolution.  Then, the conserved quantity $\Lambda$ is
\begin{align}
    \Lambda=-\frac{1}{2\pi}\frac{dI}{d\psi}.
\end{align}
For example, the aligned dipole pulsar has $I(\psi)$ given to an excellent approximation by \cite{Gralla16b}
\begin{align}
    I=\pm2\pi\Omega\psi\br{2-\frac{\psi}{\psi_o}-\frac{1}{5}\pa{\frac{\psi}{\psi_o}}^3},\quad
    \psi<\psi_o.
\end{align}
The $+$ refers to field lines that asymptote to the northern hemisphere, while the $-$ refers to field lines that asymptote to the southern hemisphere.  The expression holds for $\psi<\psi_o$, with $\psi_o$ given by
\begin{align}
    \psi_o=\sqrt{\frac{3}{2}}\mu\Omega.
\end{align}
Thus, in this case, the conserved quantity is given by
\begin{align}
    \Lambda=\mp2\Omega\br{1-\frac{\psi}{\psi_o}-\frac{2}{5}\pa{\frac{\psi}{\psi_o}}^3},
\end{align}
where the upper/lower sign refers to the northern/southern flow.

\section{Simulations and fit}
\label{app:Fit}

We carried out a number of time-dependent 3D simulations of oblique pulsar magnetospheres in flat spacetime in the force-free approximation with the code by \citet{Spitkovsky06} in a Cartesian grid.  We consider a perfectly conducting star of radius $R_\star$ with a magnetic dipole field of dipole moment $\mu$ that makes an angle $\iota$ with the rotational axis.  The star rotates with angular velocity $\Omega$.  We performed simulations for a range of inclination angles $\iota=\cu{0^\circ,30^\circ,60^\circ,90^\circ}$ and rotation $\epsilon=\cu{0.4,0.2,0.1,0.067}$.  The stellar radius is resolved by 40 computational cells in all simulations.  We find that the conserved quantity is essentially converged, in that $\Lambda(\alpha,\beta)/\epsilon$ changes very little between $\epsilon=0.1$ and $\epsilon=0.067$.  The rescaled polar cap size $\alpha_0/\epsilon$ shows a small increase with decreasing $\epsilon$, tending toward the reported value \eqref{eq:LastOpenFieldLine}.  The non-uniformity of the asymptotic magnetic field \cite{Tchekhovskoy16}, which is linked to the functional form of $\Lambda(\alpha,\beta)$, does not depend on $\epsilon$.  The increase of the polar cap size (or the value of the open magnetic flux) with decreasing $\epsilon$ leads to an increase in the spin-down power \cite{Philippov14}.  We perform the fit for $\epsilon=0.067$.  Figure~\ref{fig:Fit} shows the results.\hfill\includegraphics[width=.01\textwidth]{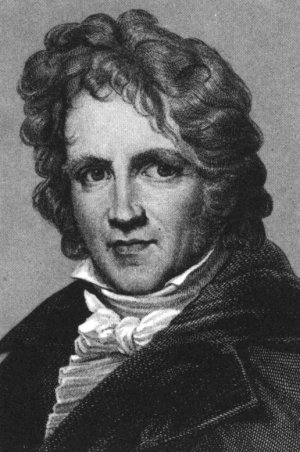}

\bibliographystyle{apsrev4-1}
\bibliography{InclinedPulsarMagnetospheres}

\end{document}